# How does the Shift-insertion sort behave when the sorting elements follow a Normal distribution?

**Mita Pal, Soubhik Chakraborty, N.C. Mahanti**
**Department of Applied Mathematics,**
**B.I.T, Mesra, Ranchi-835215, India**

**ABSTRACT:** The present paper examines the behavior of Shift-insertion sort (insertion sort with shifting) for normal distribution inputs and is in continuation of our earlier work on this new algorithm for discrete distribution inputs, namely, negative binomial. Shift insertion sort is found more sensitive for main effects but not for all interaction effects compared to conventional insertion sort.

## Introduction

The present paper examines the behavior of Shift-insertion sort (insertion sort with shifting) for normal distribution inputs and is in continuation of our earlier work on this new algorithm for discrete distribution inputs, namely, negative binomial [PCM10a]. The focus is on parameterized complexity, using factorial experiments, where the n observations to be sorted come from normal population N (m, s). To investigate the individual effects of number of sorting elements (n), normal distribution parameters and also their joint effects, a 3-cube factorial experiment is conducted with three levels of each of the factors n, m and s. Here m and s stand for "mu" and "sigma", the two population parameters representing mean and standard deviation for normal distribution respectively. For brevity in typing, we have avoided the symbols μ (mu) and σ (sigma) that we use generally.

Referring to [Knu00] for a comprehensive literature on sorting algorithms and to [Dob10] for insertion sort, we briefly distinguish between insertion sort and shift insertion sort first before moving to the main experimental findings. The idea of insertion sort is based on considering every element in an array, one at a time, and comparing it with preceding elements that are already sorted. So the algorithm finds the correct position in the prefix and inserts considered element into this position to make the sub-array sorted. That is, we insert the $i^{th}$ element A[i] into its rightful place on the $i^{th}$ pass among A[1], A[2], …., A[i-1], which were previously placed in sorted order. After doing this insertion, the records occupying A[1],…,A[i] are in sorted order. The next section describes our shift-insertion sort, a modified version of insertion sort involving shifting, suitable for sorting short lists only but is faster than bubble sort and selection sort. It is simpler to program than quick sort. Shift insertion sort was introduced first in [PCM10b] where it was found to be faster than the conventional insertion sort for uniform inputs. However Shift-insertion may not be faster than all versions of insertion sort.

The next section describes our shift-insertion sort.

## 1 Shift-Insertion sort (insertion sort with shifting)

Step 1: Repeat steps 2 to 7 for j = 1 to n.
Step 2: Repeat steps 3 to 7 for i = 0 to j
Step 3:  Compare A[i] and A[j]. If A[i]>A[j]
            then repeat step 4 to 7.
Step 4: Set temp:=A[j].
Step 5:  Repeat steps 6   for k = j to i.
Step 6: Set A[k]:=A[k-1].
Step 7:  Set A[i]:=temp.

## 2. Empirical results

Assume the sorting elements follow a Normal N (m , s) distribution. Here n independent N (m, s) variates are obtained by simulation (using Box Muller transformation [KG80]) and kept in an array and sorted. Tables 1 and 2 give the data for factorial experiments to accomplish our study on parameterized complexity.  In what follows, insertion sort will refer to the conventional insertion sort.

**Table 1. Data for $3^3$ factorial experiment for insertion sort**

Shift-insertion sort times in seconds for normal (m, s) distribution input for various n (5000, 7000, 9000), s (800, 1200, 1600) and m (500, 1000, 1500)

s = 800

| n | m=500 | m=1000 | m=1500 |
|---|---|---|---|
| 5000 | 0.058 | 0.060 | 0.052 |
| 7000 | 0.113 | 0.102 | 0.102 |
| 9000 | 0.179 | 0.173 | 0.170 |

s = 1200

| n | m=500 | m=1000 | m=1500 |
|---|---|---|---|
| 5000 | 0.055 | 0.049 | 0.055 |
| 7000 | 0.113 | 0.107 | 0.104 |
| 9000 | 0.170 | 0.181 | 0.169 |

s = 1600

| n | m=500 | m=1000 | m=1500 |
|---|---|---|---|
| 5000 | 0.055 | 0.058 | 0.058 |
| 7000 | 0.107 | 0.113 | 0.107 |
| 9000 | 0.170 | 0.173 | 0.181 |

**Table 2. Data for $3^3$ factorial experiment for Shift-insertion sort**

Shift-insertion sort times in seconds for Normal N(m , s) distribution input for various n (5000, 7000, 9000), s (800, 1200, 1600) and m (500, 1000, 1500)

s = 800

| n | m=500 | m=1000 | m=1500 |
|---|---|---|---|
| 5000 | 0.041 | 0.047 | 0.044 |
| 7000 | 0.093 | 0.093 | 0.091 |
| 9000 | 0.148 | 0.159 | 0.148 |

s = 1200

| n | m=500 | m=1000 | m=1500 |
|---|---|---|---|
| 5000 | 0.044 | 0.047 | 0.047 |
| 7000 | 0.098 | 0.096 | 0.096 |
| 9000 | 0.151 | 0.153 | 0.153 |

s = 1600

| n | m=500 | m=1000 | m=1500 |
|---|---|---|---|
| 5000 | 0.052 | 0.044 | 0.047 |
| 7000 | 0.091 | 0.102 | 0.085 |
| 9000 | 0.157 | 0.157 | 0.153 |

Table 3 gives the results using MINITAB statistical package version 15.

**Table 3. Results of $3^3$ factorial experiment on insertion sort**

**General Linear Model: y versus n, s, m**

| Factor | Type | Levels | Values |
|---|---|---|---|
| n | Fixed | 3 | 0, 1, 2 |
| s | Fixed | 3 | 0, 1, 2 |
| m | Fixed | 3 | 0, 1, 2 |

**Analysis of Variance for y, using Adjusted SS for Tests**

| Source | DF | Seq SS | Adj SS | Adj MS | F | P |
|---|---|---|---|---|---|---|
| n | 2 | 0.1901147 | 0.1901147 | 0.0950574 | 11457.81 | 0.000 |
| s | 2 | 0.0000734 | 0.0000734 | 0.0000367 | 4.42 | 0.017 |
| m | 2 | 0.0000927 | 0.0000927 | 0.0000463 | 5.58 | 0.006 |
| n*s | 4 | 0.0000210 | 0.0000210 | 0.0000052 | 0.63 | 0.642 |
| n*m | 4 | 0.0000888 | 0.0000888 | 0.0000222 | 2.68 | 0.041 |
| s*m | 4 | 0.0001779 | 0.0001779 | 0.0000445 | 5.36 | 0.001 |
| n*s*m | 8 | 0.0002484 | 0.0002484 | 0.0000310 | 3.74 | 0.002 |
| Error | 54 | 0.0004480 | 0.0004480 | 0.0000083 | | |
| Total | 80 | 0.1912649 | | | | |

S = 0.00288033   R-Sq = 99.77%   R-Sq(adj) = 99.65%

**Table 4. Results of $3^3$ factorial experiment on shift-insertion sort**

**General Linear Model: y versus n, s, m**

| Factor | Type | Levels | Values |
|---|---|---|---|
| n | Fixed | 3 | 0, 1, 2 |
| s | Fixed | 3 | 0, 1, 2 |
| m | Fixed | 3 | 0, 1, 2 |

**Analysis of Variance for y, using Adjusted SS for Tests**

| Source | DF | Seq SS | Adj SS | Adj MS | F | P |
|---|---|---|---|---|---|---|
| n | 2 | 0.1523912 | 0.1523912 | 0.0761956 | 11868.93 | 0.000 |
| s | 2 | 0.0001352 | 0.0001352 | 0.0000676 | 10.53 | 0.000 |
| m | 2 | 0.0001962 | 0.0001962 | 0.0000981 | 15.28 | 0.000 |
| n*s | 4 | 0.0002306 | 0.0002306 | 0.0000576 | 8.98 | 0.000 |
| n*m | 4 | 0.0000618 | 0.0000618 | 0.0000154 | 2.40 | 0.061 |
| s*m | 4 | 0.0001049 | 0.0001049 | 0.0000262 | 4.08 | 0.006 |
| n*s*m | 8 | 0.0002109 | 0.0002109 | 0.0000264 | 4.11 | 0.001 |
| Error | 54 | 0.0003467 | 0.0003467 | 0.0000064 | | |
| Total | 80 | 0.1536774 | | | | |

S = 0.00253372   R-Sq = 99.77%   R-Sq(adj) = 99.67%

## 2. Discussion

Shift-insertion sort is highly affected by the main effects n, m and s than insertion sort. When we consider the interaction effects, interestingly we find that all interactions are more significant in the case of Shift-insertion sort compared to insertion sort except n*m and s*m. Also, In the case of insertion sort the interaction of n*s is quite insignificant.

Comparing the numerical values of calculated variance ratio F from tables 3 and 4, we have another interesting finding on the sensitivity of these two algorithms towards the seven factors (three main effects and four interactions) as summarized in table 5. Note that "more" means comparatively more and "less" means comparatively less in table 5.

**Table 5. Summary of Parameterized Complexity**
**(sensitivity to main and interaction effects)**

| Factor | Shift-insertion sort | Insertion Sort |
|--------|---------------------|----------------|
| n | More | Slightly Less |
| s | More | Less |
| m | More | Less |
| n*s | More | Very less |
| n*m | Insignificant at 5% level | Significant at 5% level |
| s*m | Slightly Less | More |
| n*s*m | More | Less |

It is evident from table 5 that shift insertion sort is more sensitive to main effects as compared to insertion sort and it is also sensitive to interaction effects than insertion sort except for the interaction n*m and s*m for which insertion sort continues to be slightly more sensitive. In particular, the less sensitivity to n for Shift-insertion sort makes it faster than insertion sort even for non-uniform inputs such as normal. This concludes our discussion.

## Conclusion and suggestions for future work

Three-cube factorial experiments conducted on Shift-insertion sort and insertion sort reveal that for certain algorithms such as sorting, the parameters of the input distribution singularly as well as interactively are important factors, besides the size of input, for evaluating time complexity more precisely. While, our results will definitely pose an intellectual challenge for the theoretical analysts, we do emphasize here that cheap and efficient prediction (see [CMP09] and [S+89]) is the objective in computer experiments such as the ones conducted here. A computer experiment is a series of runs of a code for various inputs.